# Magnetic field dependence of the magnetic phase separation in $Pr_{1-x}Ca_xMnO_3$ manganites studied by small-angle neutron scattering

Damien Saurel[1,2], Annie Brûlet[2], André Heinemann[3], Christine Martin[1], Silvana Mercone[1], Charles Simon[1]

[1]Laboratoire CRISMAT, UMR 6508, ENSICAEN-CNRS, 6 Boulevard du Maréchal Juin, 14050 Caen, France

[2]Laboratoire Léon Brillouin (UMR 12 CEA-CNRS), CEA Saclay, 91191 Gif sur Yvette, France

[3]Hahn Meissner Institute, Glienicker Straße 100, D-14109 Berlin, Germany.

**Abstract**

Transport properties of manganese oxides suggest that their colossal magnetoresistance (CMR) is due to percolation between ferromagnetic metallic (FM) clusters in an antiferromagnetic insulating (AFI) matrix. We have studied small-angle neutron scattering under applied magnetic field in CMR $Pr_{1-x}Ca_xMnO_3$ crystals for x around 0.33. Quantitative analysis of the small-angle magnetic neutron scattering shows that the magnetic heterogeneities take place at different scales. At the mesoscopic scale (200nm), the inhomogeneities correspond to the percolation of the conducting ferromagnetic phase into the insulating phases. It is at the origin of the colossal magnetoresistance of the compound. The other inhomogeneities are nanoscopic: inside the antiferromagnetic phase (AFI), there exist small ferromagnetic clusters. Inside the ferromagnetic phase which exists in absence of magnetic field in some compounds and is in fact insulating (FI), there also exist small non ferromagnetic objects. No evolution of this nanostructure is observed when the magnetic field is applied. The existence of such nanoscale objects is discussed in relation to the cationic disorder of these compounds.

PACS

| | |
|---|---|
| 75.47.Gk | Colossal magnetoresistance |
| 75.47.Lx | Manganites |
| 75.25.+z | Spin arrangements |

**I. Introduction**

The mixed valence perovskites $A_{1-x}B_xMnO_3$ attract a considerable attention for their spectacular colossal magnetoresistance properties [1]. The exchange rules established in the 50's by Zener and Goodenough permit a first understanding of the complex mechanisms driving the transport properties of such compounds. Essentially it deals with a competition between an antiferromagnetic charge ordered (AF-CO) ground state according to the super-exchange rules and a ferromagnetic (F) coupling associated with a delocalization of the charge carriers. But these theories, and few years later the de Gennes canting theory, cannot explain the colossal and abrupt fall of resistivity observed in the most spectacular colossal magnetoresistive (CMR) compounds [2, 3]. A new theory to explain this effect is a percolation between ferromagnetic clusters in a magnetic phase separated system [2]. This idea is supported by theoretical calculations and recent experimental measurements [4, 5]. The phase separation, often discussed by theoretical groups [6, 7] and experimentally observed [8, 9], is a micrometric one with no peculiar geometry. But few theoretical considerations taking into account Coulomb repulsion between the ferromagnetic clusters predicts a topology of these later as a "mist" of nanometric droplets rather than large micrometric regions [10]. In some particular cases, it predicts that these clusters can have the shape of thin elongated objects, for instance due to the hopping of conducting electrons [11], or of anisotropic objects [12, 13] due to anisotropic interactions with the lattice and consequently due to interfacial energy between F and AF.

Among the manganite compounds, the $Pr_{1-x}Ca_xMnO_3$ series is of prime interest, because Pr and Ca have the same size and hence minimize the cationic size mismatch. Depending on x, the $Pr_{1-x}Ca_xMnO_3$ compounds show a large variety of states [14]. The CMR effect is observed in compounds of a narrow range of composition between x=0.3 and x=0.4, where coexistence of two micrometric magnetic phases in absence of magnetic field exists [18]: a ferromagnetic one that we will show here to be in fact insulating (FI) and an antiferromagnetic insulating phase (AFI) are observed. In presence of magnetic field, these phases transform into a ferromagnetic metallic phase (FM). Compositions close to x=0.33 have been widely studied by magnetoresistance, magnetization, neutron diffraction and specific heat measurements [15, 16]. Some results have been obtained by magnetic small angle neutron scattering MSANS [4, 5, 17, 18, 19, 20], which is a powerful technique to study the (micro)phase separation in the range 1-50 nm between ferromagnetic and non-ferromagnetic domains due to large magnetic contrast between them. This experimental technique was first used in manganites by de Teresa et al. [17] and more recently by Radaelli et al. [18] and Viret et al. [19]. A quantitative analysis of the scattered intensity was presented in recent papers [4, 5], showing that at low temperature (30K) and in absence of magnetic field, the structure of a $Pr_{0.67}Ca_{0.33}MnO_3$ single crystal was quite peculiar looking like a "red cabbage" with randomly oriented F sheets of about 1 nm of average thickness.

The three $Pr_{1-x}Ca_xMnO_3$ single crystals which are studied here display large CMR effects, and contain three different FI/AFI ratio. We performed a small angle neutron scattering study combined to magnetization measurements in order to get information on the evolution of the size and the shape of this phase separation under an applied magnetic field.

## II. Experimental

We have measured by classical extraction magnetometry the magnetization of three $Pr_{1-x}Ca_xMnO_3$ samples with x in the range 0.3-0.4, cut out of crystals grown using the floating zone method, as function of the magnetic field in a Quantum Design PPMS magnetometer up to 9T (zero field cooling procedure). Figure 1 presents the corresponding results showing the phase transformation induced by the application of the magnetic field at 30K. Conductivity measurements were also performed at 30K by the standard four-probe technique at fields up to 9T (figure 2). The corresponding percolation thresholds are indicated by arrows on figures 1 and 2. Typical magnetic fields of 3 to 6 T are necessary to transform the insulating phases into metallic ferromagnetic (FM) one. In order to determine the ferromagnetic fractions, we have used the method which was validated in reference 5: the extrapolation, at H=0, of the constant slope of the magnetization curve –as shown Fig. 1 by the dotted line- leads to a magnetization value which corresponds to the magnetic moment refined from neutron diffraction data [4]. The corresponding ferromagnetic (FI) percentages were found to be: 30% (+/- 3%), 6% (+/- 0.5%) and 0% (+/- 0.3%) for the three compounds x=0.30, x=0.34 and x=0.37, respectively. The remaining part of the samples is antiferromagnetic insulating (AFI). In the following, the samples will be labeled by their ferromagnetic fraction $x_F$ = 30%, 6% and 0%. Since the percentage of the ferromagnetic phase is 30% in one of the sample which is insulating, it is strongly suggested that the ferromagnetic phase is insulating, as this is the case in the x=0.2 region of the phase diagram [14]. The exact nature of this insulating phase is probably slightly different from 0.2 to 0.3: this would deserve a more detailed study. Another point has to be underlined: the role of the Pr moments is rather complex and is not studied here. Neutrons studies on the composition x=0.33 revealed a contribution of about 0.4 to 0.5μB to the total magnetization, appearing under 30K [28].

Small angle neutron scattering were performed on PAXY spectrometer of LLB and on V4 spectrometer in Hahn-Meissner Institute. Different experimental configurations were used to perform scattering measurements in the Q range from 0.1 nm$^{-1}$ to 2 nm$^{-1}$. The samples (typically 10x6x3mm$^3$) were introduced in a cryostat with aluminum windows. No precise orientation of the crystal was chosen with respect to the neutron beam since no anisotropy was observed in small angle neutron scattering without applied magnetic field. The magnetic field was applied horizontally, perpendicular to the neutron beam. Few parasitic reflections were removed before data treatment. In order to subtract the background signal, an empty cell was measured. The calibration of the spectrometer was performed with a Plexiglas sample; the scattering function was determined in absolute units (cm$^{-1}$) [21]. We have systematically neglected the inelastic spin wave corrections. Without magnetic field, an isotropic pattern is observed (figure 3a). In this case, we used the software REGISO (from LLB) to integrate the angular intensity. By applying magnetic field, the orientation of the magnetic moments with respect to the direction of the field gives a typical anisotropic picture (figure 3b) [22]. The total scattering can be separated into an isotropic $I_A$ and anisotropic $I_B \sin^2\alpha$ parts [22]:

$$I(Q) = I_A(Q) + I_B(Q)\sin^2\alpha \qquad\qquad (1)$$

where $\alpha$ is the angle between the scattering vector Q and the direction of the magnetic field. The quantities $I_A(q)$ and $I_B(q)$ can be measured precisely from the analysis of the whole 2D pattern according to equation (1). We have used the BERSANS software from HMI. On the V4 spectrometer, a polarization analysis allows to determine the magnetic-nuclear cross term [22]. On the single crystal $x_F=30\%$, no measurable signal was obtained from the difference between the intensities of the two polarization states: the nuclear signal is already quite small and thus the cross term is even smaller. The arithmetic mean of the two polarization states intensities corresponds to the intensity of a non –polarized beam given by equation (1). The purely magnetic $I_B$ component is the intensity scattered by the magnetism oriented by the magnetic field.

### III. Results and discussion

Figure 4 presents absolute values of the scattering part $I_B(Q)$ of the $x_F=0.30$ compound, typically obtained when applying magnetic field above 2T at low temperature (30K). One has observed qualitatively the same $I_B(Q)$ curves in the $x_F=0.06$ single crystal - not shown. The pure $x_F=0$ AFI sample was not studied in this Q range. One can distinguish on this figure that there exist two components on these curves, one at low Qs which varies in Q$^{-4}$ (it does not exist at 2T), and a second component at larger Qs varies typically as $Q^{-2}$. Since the magnetic field dependences of these two components are very different, we will discuss them independently.

***Analysis and evolution of the $I_B$ signal at low Qs :***

Let us start our discussion with the $Q^{-4}$ component which is observed at small Qs. Such a signal is characteristic of sharp interfaces [24] of mesoscopic domains of few hundreds nm large. In order to extract typical sizes, it was necessary to measure the scattering intensity in absolute units. The magnetic field evolutions of these $Q^{-4}$ signal in the $x_F$=0.30 and $x_F$=0.06 single crystals are represented on figure 5a. Note that this component is proportional to the specific area of the interfaces (i.e. the total interfaces area divided by the sample volume). At low magnetic fields, there is no $Q^{-4}$ component. This signal appears at fields sufficient to induce the magnetic transformations observed on the magnetization curves of figure 1 (larger than 2T and 3T in the $x_F$=0.06 and $x_F$=0.30 compounds, respectively). Then, $I_B Q^4$ increases with the magnetic field (figure 5a), goes through a maximum when the phase separation is maximum, and decreases at higher magnetic field. As shown on the pictures of figure 5b, the appearance of this small angle signal corresponds to the nucleation of the FM phase, (the corresponding size of these domains is typically 200 nm assuming a complete contrast between FM and AFI) and its decreases to collapse (fig 5b). The corresponding magnetic field evolutions of the sizes of the FM and AFI phases are presented on figure 6. This phase separation observed here at large scale is at the origin of the CMR effect. Since the domains involved in this scattering are quite large compared to our experimental resolution, it was not possible to be more precise about their shapes. However, it is one of the first time that a bulk technique is able to follow the size evolution of the percolating domains, confirming in a very elegant manner the existence of percolation in this system.

***Analysis of the $I_B$ signal observed at large Qs.***

The second component of the $I_B$ scattering observed at higher Q range corresponds to the scattering by the nanometric domains. Let us present on figure 7 the results obtained on the $x_F$=0.30 composition at low temperature (2K) and moderate magnetic field (2T), enough to orientate the magnetic moments, but not enough to transform the AFI phase into a FM one. The absolute values of the scattering parts $I_B(Q)$. A good agreement is obtained between the different configurations (wavelengths and detector distances) which were necessary to record the scattering intensities in the 0.06 to 2 nm$^{-1}$ range of wave vectors. The resolution reaches $10^{-2}$ cm$^{-1}$. Thanks to these conditions, it is possible to observe three regimes of the magnetic small angle scattering: a plateau at low Q, a Q$^{-n}$ region in the medium Q range, with n~2, and a rapid decrease at higher Q. The transitions between these three regimes correspond to two typical sizes of the system, $R_G$ the Guinier radius of gyration of scattering objects and r their smallest characteristic size. A good approximation of the scattering intensity in the high Q regime $QR_G >> 1$ is given by a Guinier function [23]:

$$I(Q) = KQ^{-n} \exp(-Q^2 r^2/3) \qquad (2)$$

In this intermediate scattering range defined by Porod [24], the scattered intensity obeys to a power law Q$^{-n}$, which depends on the shape or on correlation of the scattering objects [23, 24]. For instance, a Q$^{-2}$ dependence is characteristic of flat objects, a Q$^{-1}$ law is attributed to long ones. This intermediate Q range extends from Q~2-3$R_G^{-1}$ to Q~$r^{-1}$ where r stands for the thinnest size of the scattering objects (thickness $t = 2r$ for layers, mean radius $\bar{r} = \sqrt{4/3}r$ of the cross section for strings). The factor K depends on the phase fraction, the scattering contrast and the form factor of the scattering objects. The plateau at low Q indicates the largest typical size of the system, above which the sample appears as homogeneous (typically 10 nm). In order to determine the shape of the objects, we have analyzed the experimental data on figure 7 with different models already proposed in the literature:

(i) The Lorentzian shape was used by De Teresa et al. [17]:

$$I(Q) \propto \frac{\xi^3}{1 + Q^2\xi^2} \qquad (3)$$

This scattering function does not take into account a finite lower limit of the size of the scattering objects. For this reason the fit is not good for data at large Q. Let us note that no phase fraction can be deduced from the fit with equation (3).

(ii) The "red cabbage" structure was proposed by Simon et al. [4] in order to explain the Q$^{-2}$ signal observed in a wide Q range. Moreover, in order to take into account the presence of a Guinier regime and of a plateau on the experimental curve, it is necessary to introduce a mean lateral size (2R) of the cabbage structure. The scattering intensity of randomly oriented disks is:

$$I(Q) = \phi(\Delta\rho_m)^2 . 2\pi . t . Q^{-2}\left[1 - \frac{J_1(2QR)}{QR}\right]\exp(-Q^2 t^2/12) \qquad (4)$$

in which t is the thickness of disks, and J$_1$(x) the first Bessel function. $\phi$ is the phase fraction of scattering objects (assumed here much smaller than 1) and $(\Delta\rho_m)^2$ the magnetic contrast. Here we have assumed a magnetic contrast between F and NF regions, with ideal magnetization of 3.8μ_B/f.u. (see figure 1) and zero, respectively. $(\Delta\rho_m)^2 = (\delta.\Delta M)^2$ (where $\delta$ is the scattering amplitude of one $\mu_B$ moment of $0.27\times10^{-12}$ cm/$\mu_B$ divided by the cell unit volume of $0.57\times10^{-22}$ cm$^3$). $(\Delta\rho_m)^2 = 3.24 \ 10^{20}$ cm$^{-4}$. Despite all these modifications, the fit with equation (4) is rather poor: the exponent in the intermediate Q range is around 1.7, thus clearly smaller than 2 expected from equation (4). The fit gives a phase fraction $\phi$ of about 5%.

(iii) As proposed by Viret et al. [19], a model of self avoiding chains better explains the exponent value 1.7 observed. The expression of scattering function of such chains [25] taking also into account a finite lateral size r of chains is given in table 1. The quite good fit gives a phase fraction $\phi$ of about 0.03. However, this model suggests that the physics of the magnetic transformation is related to a one dimensional process [19]. For this reason, we have tried to fit the data with a model which does not contain this physics and is three dimensional.

(iv) The curves can indeed be fitted with an empirical expression of scattering by aggregates of spherical particles of radius $r_0$, correlated with a correlation function g(r) = K r$^{-1}$ exp(-r/ξ) (ξ is the correlation length) [26].

$$I(Q) = \phi \, (\Delta\rho_m)^2 \frac{4}{3}\pi r_0^3 \left[ 1 + \frac{8\pi K \xi^2}{\left[1+Q^2\xi^2\right]} \right] \exp(-Q^2 r_0^2/5) \qquad (5)$$

In this case also, we can obtain a good fit.

As a matter of fact, with these different geometrical models (red cabbage, self avoiding chains, lorentzian clusters, correlated assembly), one get very similar parameters (largest size: R, $R_G$ or $\xi$; thinnest size t or $r_0$; and phase fraction $\phi$). Moreover, models (iii) and (iv) fit as well the data. It indicates that it will be very difficult to be very precise concerning the structure of this nanoscopic phase separation, and we can use indifferently models (iii) and (iv) to analyze our data. The typical value obtained for R/$R_G$/$\xi$ and t/$r_0$ is around 6 nm and 1.3 nm, respectively. The value of t/$r_0$ corresponds to a single unit cell and the R/$R_G$/$\xi$ value is also nanoscopic. The corresponding phase fraction $\phi$ of the sample $x_F$=30% is very small, typically 0.04. Let us note that this very small value of the phase fraction (4%), compared to the ferromagnetic insulating (FI) phase fraction determined from magnetization (30%), indicates that the signal of this nanoscopic phase separation can hardly come from a dispersion of the FI in the AFI one. It is consistent with observations of P. G. Radaelli and collaborators [18] which have shown that the FI and AFI phases are separated at the micrometric scale. The transformation by application of a stronger magnetic field and the comparison of this effect for different compositions may help to understand the origin of this signal.

### *Evolution of $I_B(Q)$ for different compositions*

Let us then investigate the evolution of the scattering curves versus sample composition (i.e. F fraction). The $I_B(Q)$ functions are very similar –not shown- in this range of wave vectors. The characteristic sizes of scattering objects at the origin of this signal are not composition dependent. The only adjustable parameter is the phase fraction $\phi$.

On figure 8, we have reported the value of $I_B$ at $1nm^{-1}$ at 30K, as a function of the ferromagnetic fraction ($x_F$=0.30, 0.06 and 0) obtained from the magnetization data (figure 1), at a magnetic field just sufficient to oriented the magnetic domains. One can see clearly that $I_B$ increases proportionally to the percentage of the FI phase, indicating that FI is the phase which contains this nanoscale phase separation. Non ferromagnetic (NF) nanoscopic objects in the FI matrix are thus the scattering objects responsible of this MSANS.

### *Evolution of $I_B(Q)$ by application of magnetic field*

On figure 9, we have reported typical Q dependence of the $I_B$ MSANS for different magnetic fields at 10K for the sample $x_F$=0.30. As already observed [5, 19], the $I_B$ signal is decreasing without any change of its shape. It means that the decrease of $I_B$ is not due to a change of the size and the shape of the scattering objects (not a change of their form factor) contrary to what we have previously published on the basis of a smaller Q range data [5]. $I_B$ is thus simply proportional to the phase fraction occupied by these nanometric objects and their magnetic contrast with the rest of the sample. Let us now observe, figure 10b, the evolution of this signal in the $x_F$=30% and $x_F$=6% single crystals measured at Q=1 $nm^{-1}$ versus the applied magnetic field, at 30K.

In the regime of constant slope of the magnetization curves (reported figure 10), $I_B$ is constant. It confirms that these objects are NF inhomogeneities in the low susceptibility FI phase rather than in the high susceptibility AFI phase. At higher magnetic fields (larger than 3.5T and 2T in the $x_F$=0.30 and $x_F$=0.06 compounds respectively), the intensity decreases when the transformation of the AFI phase is observed on M(B) curves (figure 10). It means that FI and AFI phases transform simultaneously, and that there is no nucleation of NF nanometric objects.

The interpretation presented here is that the decrease of the MSANS intensity is only due to the disappearance of the FI phase as the magnetic field is applied. This signal is not directly related to the percolation of the FM phase and to the CMR properties, but its evolution reveals the transformation of the FI phase. This transformation occurs at higher scales, probably by nucleation of large FM regions in the FI phase. Since there is no contrast between FI and FM regions, this hypothesis cannot be confirmed.

### *Evolution of the non oriented part of the SANS: $I_A$*

Let us discuss now the evolution of the isotropic signal $I_A$. On figure 11, we have reported typical data at 30K for different magnetic fields (here for the composition $x_F=0.030$, but one observes qualitatively the same curves in the compositions $x_F=0$ and $x_F=0.06$). One can clearly see that the signal depends on the magnetic field, suggesting that it contains also a significant magnetic contribution. This result was not expected as such high fields (above 2T). There are clearly two components in this signal. At low Q, the $Q^{-4}$ signal is not depending on the magnetic field. It comes from the crystal edges (the intensity is compatible with this interpretation). At higher Q, the signal varies in $Q^{-1}$, suggesting scattering objects such as statistically oriented 1D rods [24] :

$$I(Q) = \phi \frac{2}{3}\pi^2 R^2 (\Delta\rho_m)^2 Q^{-1} \exp(-Q^2 R^2/4) \qquad (6)$$

This formula was established for $1/R \gg Q \gg 1/L$, in the case of low concentrated F rods of length L and radius R, embedded in an antiferromagnetic matrix. The 2/3 factor in equation (6) is to average the scattering of non-oriented magnetic objects. We have reported on figure 12 the evolution of the $I_A$ signal at high Qs ($1nm^{-1}$) for the $x_F=0.30$ and $x_F=0$ compounds. This signal decreases as magnetic field increases. Since no $Q^{-1}$ dependent signal was observed in the $I_B$ part of the signal, this decrease does not correspond to a concomitant rotation of the spins induced by the magnetic field. It can hardly arise from the FI phase, the corresponding moments being observed aligned by a magnetic field of 2T. It is more probable that this signal comes from small ferromagnetic elongated and rigid objects which exist in the AFI phase. The decrease of the intensity could be due to the contrast or to the number of scattering objects. The application of the magnetic field on the AFI phase induces a magnetization which indeed decreases the contrast, but would have produce an increase of the $I_B$ signal which is not observed. It is thus more probable that the volume of the AFI phase and then the number of scattering objects are decreasing as the magnetic field is applied.

Three kinds of interactions can be invoked as origin of the rod-like shape of the F clusters: exchange, elastic and magnetic dipolar interactions. The hopping electrons in the AF ordered environment, as discussed by Kagan et al. [11] and Viret et al. [19], generates F conducting filaments in the AF matrix. The spin of such moving electrons aligns the spins parallel to the direction of the filament by a generalized Hund's rule. In this quantum effect, the F ordering in the AF matrix occurs without changing the lattice distortion associated to the AFI state. Thus, the interfaces of this 1D nanometric filamentary structure cost no elastic distortion energy to the system. The number and the length of such F hopping strings only depend on the competition between double exchange and super exchange. In this case, the filaments can be randomly oriented, but their rigidity evidenced by the observed $Q^{-1}$ power law, as well as the fact that their magnetization is not oriented by a magnetic field of 2T, needs to take into account other interactions. In a more simple way, these defects can also be some discommensurations in the AFI structure.

### *Discussion*

The presence of small clusters in both the FI and AFI phases can be seen as a consequence of the presence of cationic disorder of the perovskites, where Pr and Ca are randomly distributed on the "Pr-Ca" site. If one thinks about the very small characteristic length found here (~2nm for the smaller one), it is important to point out that inside a sphere of this radius, there are only 500 "Pr-Ca" cations, thus the fluctuation of their local density can reach 4%. At this length scale of 2 nm, the composition of the x=0.30 sample for instance can be x=0.30±0.04. Since the magnetization of these compounds is strongly dependent of the x concentration (cf insert of figure 1) in the x=0.3-0.4 concentration range, these density fluctuations can be reasonably at the origin of the nanometric magnetic inhomogeneities [6] observed in both FI and AFI phases.

We have shown that FI and AFI phases transforms simultaneously. Both of them undergo a CMR transformation (i.e. associated I-M and magnetic transformation). These compounds are at the vicinity of the boundary between the domains of stability of the FI and AFI phases which extend at lower and higher x values, respectively [14]. Consequently, the occurrence of the CMR effect at these compositions can be related to the destabilization of these phases. The presence of nanometric inhomogeneities that we observe in these CMR compounds reveals that this destabilization has for consequence a strong sensibility of these phases to the intrinsic local chemical disorder.

**IV. Conclusion**

The use of the magnetic SANS technique under magnetic field, coupled to classical magnetization measurements allows us to determine the magnetic field dependence of the size and the shape of the different magnetic inhomogeneities in the phase separated $Pr_{1-x}Ca_xMnO_3$ system. The phase separation occurs at three different scales. At a large scale, there is a "macroscopic" phase separation between FI and AFI phases. At the intermediate scale, the evolution of the intensity scattered by the AFI phase at low Q reveals nucleation, growing and percolation of mesoscopic FM regions. At a smaller scale, nanometric clusters are observed in both AFI and FI phases. The latter produces a large signal in MSANS. From the Q dependence of the magnetic scattering, it is nevertheless difficult to choose between several possible structural models (self avoiding strings or correlated objects). The system presents two characteristic lengths, one about 1nm, the other one typically 6 nm. It is probably a mixing of small objects (1nm) clustered into larger aggregates (typically 6 nm).

Application of strong fields transforms these phases into a quite homogeneous FM phase, since there is no more MSANS signal. But, contrary to what was expected at the beginning of this study, the presence of nanometric clusters is not directly at the origin of the colossal magnetoresistance which is the most fascinating property of these manganites. Nevertheless, it is certainly related to the relatively small value of the critical field (6T), compared to the 30T which are necessary to transform the "ideal" AFI phase of the $Pr_{1/2}Ca_{1/2}MnO_3$, i.e. to destabilize the AFI phase. The role of the cationic disorder seems to be important in the presence of all these nanometric clusters.

**Acknowledgments**

We acknowledge L. Hervé for samples preparation,  and V. Hardy, A. Maignan, M. Hervieu, R. Retoux, F. Ott , who also for provided the cryostat, J. Teixeira, M. Hennion, F. Moussa and M. Viret for their very important support and numerous scientific discussions. D. Saurel is supported from CEA and "Région Basse Normandie".

Table 1: Scattering intensity by self avoiding strings of length L, of radius of cross section r and persistence length b (local rigidity of the string), occupying a fraction $\phi$ of the total volume. The form factor of an infinitely thin string is $P_1(Q, L, b)$ for $Qb < 3.1$ and $P_2(Q, L, b)$ for $Qb > 3.1$ [25]. The Guinier factor $\exp(-Q^2 r^2/4)$ is introduced to take into account the finite cross section of the string.

$Qb < 3.1$:

$$I_m(Q) = \phi(\Delta\rho)^2 P_1(Q,L,b)\exp(-Q^2 r^2/4)$$

$$P_1(Q,L,b) = \left\{ \begin{array}{l} \left[1 - \tanh((\sqrt{x} - 1.523)/0.1477)\right]/2\,\frac{2}{x^2}\left[x - 1 + \exp(-x)\right] \\ + \left[1 - \left[1 - \tanh((\sqrt{x} - 1.523)/0.1477)\right]/2\right]\left[ \begin{array}{l} 1.220\left(\sqrt{x}\right)^{-D} \\ + 0.4288\left(\sqrt{x}\right)^{-2D} - 1.651\left(\sqrt{x}\right)^{-3D} \end{array} \right] \\ + C(L/b)\left[\frac{4}{15} + \frac{7}{15x} - \left(\frac{11}{15} + \frac{7}{15x}\right)\exp(-x)\right]b/L \end{array} \right\}$$

$$x = R_G^2 Q^2$$

$$R_G^2 = \left[1 + \left(\frac{(L/b)}{3.12}\right)^2 + \left(\frac{(L/b)}{8.67}\right)^3\right]^{\frac{2/D - 1}{3}}\frac{Lb}{6}$$

$$C(L/b) = \begin{cases} 3.06(L/b)^{-0.44}, & L > 10b \\ \\ 1, & L \leq 10b \end{cases}$$

$Qb \geq 3.1$:

$$I_m(Q) = \phi(\Delta\rho)^2 P_2(Q,L,b)\exp(-Q^2 r^2/4)$$

$$P_2(Q,L,b) = \left[ \begin{array}{l} \dfrac{P_1(Q=3.1)\cdot 3.1^{4.12}}{(Qb)^{4.12}} - \dfrac{\pi b 3.1^{3.12}}{L(Qb)^{4.12}} + \dfrac{\pi}{QL} \\ \\ -\dfrac{4.12\cdot 3.1^{4.42}}{0.3}\left[ \begin{array}{l} P_1(Q=3.1) + \dfrac{dP_1}{dQ}(Q=3.1)\dfrac{q_0}{b4.12} \\ -\dfrac{\pi b}{3.1L}\left(1 - \dfrac{1}{4.12}\right) \end{array} \right]\left[\dfrac{3.1^{-0.3}}{(Qb)^{4.12}} + \dfrac{1}{(Qb)^{4.42}}\right] \end{array} \right]$$

Table 1

Figure 1: Magnetic field dependence of the magnetization of the three samples at 30K. In the inset, the corresponding values obtained at 1.45T and 5K (from ref. 14). The arrows in the main figure correspond to the field for which the IM transition is observed (determined from figure 2).

Figure 2: Magnetic field dependence of the conductivity of the three samples at 30K: oscillations observed for the sample $x_F$=0.06 are due to Joule effect instabilities [ 27].

Figure 3: The MSANS patterns of the $x_F$=0.30 sample observed at 30K without (a) and with an applied horizontal magnetic field of 2T (b).

Figure 4: The part $I_B$ of the MSANS obtained at 30K for the samples $x_F$ = 0.30 under 2T (just sufficient to orients the magnetic domains) and 4T (enough to induce the magnetic transformation associated to the CMR effect). The $Q^{-2}$ and $Q^{-4}$ dependences are also shown.

Figure 5: (a) The magnetic field dependence of $I_B Q^{-4}$ determined at low Q (0.06 nm$^{-1}$) is represented in absolute units for the $x_F$=0.30 and $x_F$=0.06 compounds. (b) A schematic drawing of the evolution of the field induced phase separation between FM and AFI phases is shown, corresponding to the situations labelled A, B, C and I, II, III in figure 5a..

Figure 6: Magnetic field evolution of the size of the FM (black symbols) and AFI (open symbols) domains in $x_F$=0.30 sample at 30K.

Figure 7: The $I_B$ part of the MSANS as function of the scattering vector Q for the sample $x_F$=0.03 at 2T, 2K. On the same figure, different fits to the models described in the text are shown (a) the Lorentzian function (b) disk of radius R and thickness t (c) self-avoiding chain of radius of gyration $R_g$ and radius of cross section (d) correlated cluster of size $\xi$.

Figure 8: Evolution of the $I_B$ part of the MSANS at Q=1nm$^{-1}$ and at a magnetic field just sufficient to orientate the magnetic domains (2T for $x_F$=0.30 and $x_F$=0; 1T for $x_F$=0.06) as function of the ferromagnetic fraction $x_F$.

Figure 9: The magnetic field dependence of the $I_B$ part of the MSANS versus Q, of the sample $x_F = 0.30$ at 10K.

Figure 10: The magnetic field dependence of the $I_B$ part of the MSANS at $Q=1nm^{-1}$ at 30K for the samples $x_F = 0.30$ and x=0.06. On the upper part of the figure, the corresponding evolution of the magnetization is reported for comparison.

Figure 11: The isotropic part $I_A$ of the MSANS as function of the scattering vector Q of the sample $x_F = 0.06$ at 30K for various fields 1T, 2T and 4T.

Figure 12: The isotropic part $I_A$ of the MSANS at $Q=1nm^{-1}$ as function of the magnetic field of the samples $x_F = 0.30$ and $x_F=0$, at 30K.

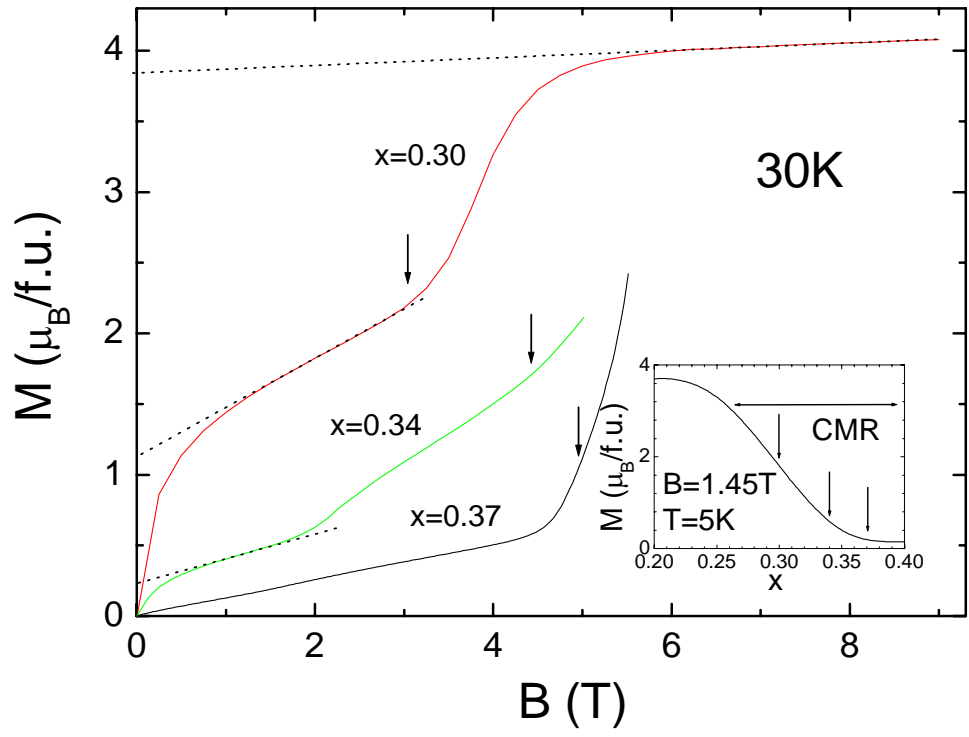

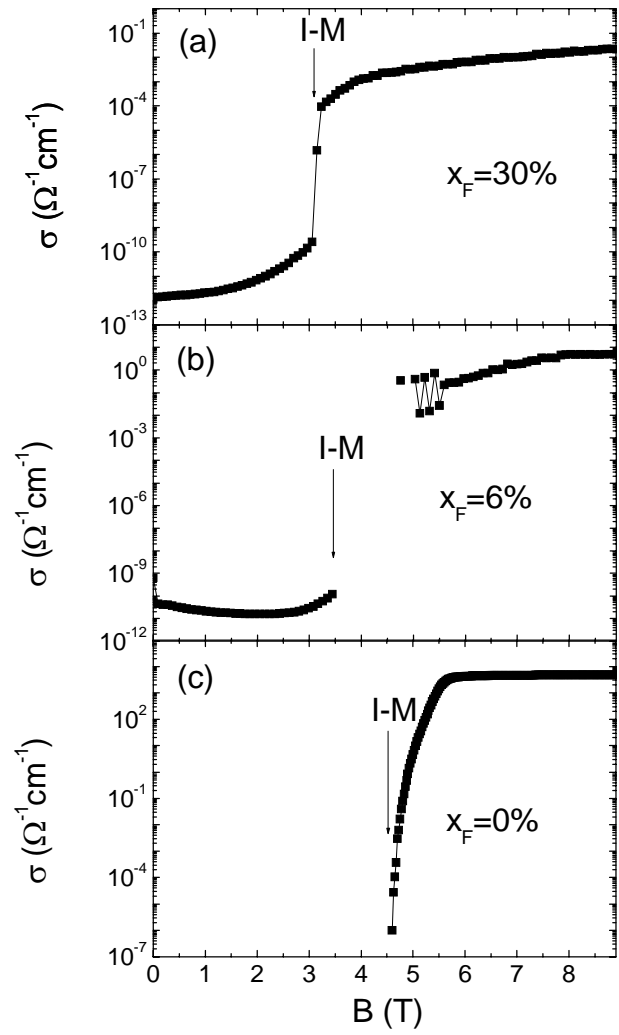

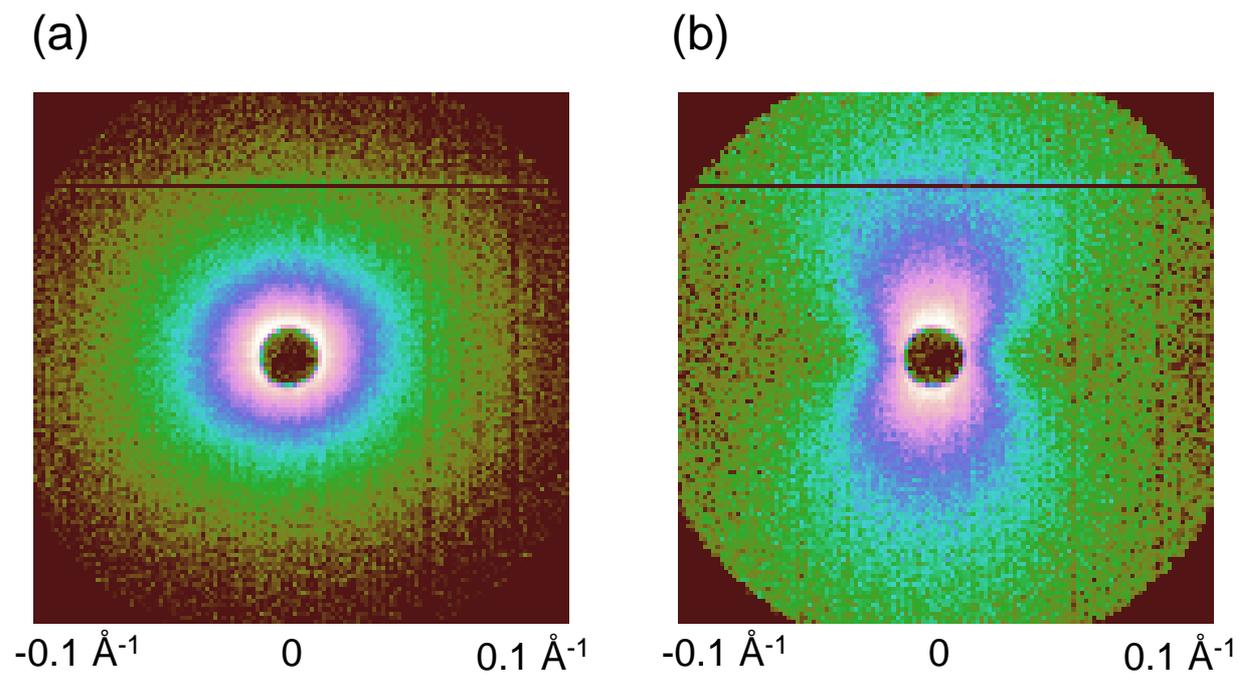

(a)

(b)

-0.1 Å⁻¹    0    0.1 Å⁻¹

-0.1 Å⁻¹    0    0.1 Å⁻¹

*Figure 3*

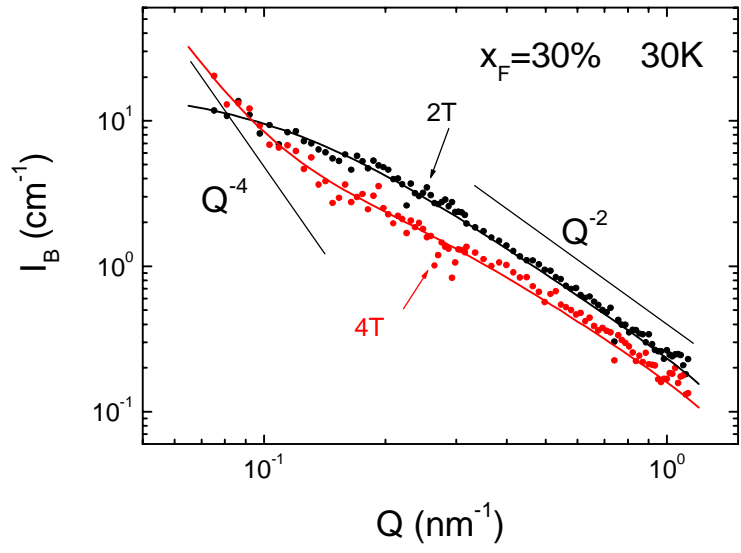

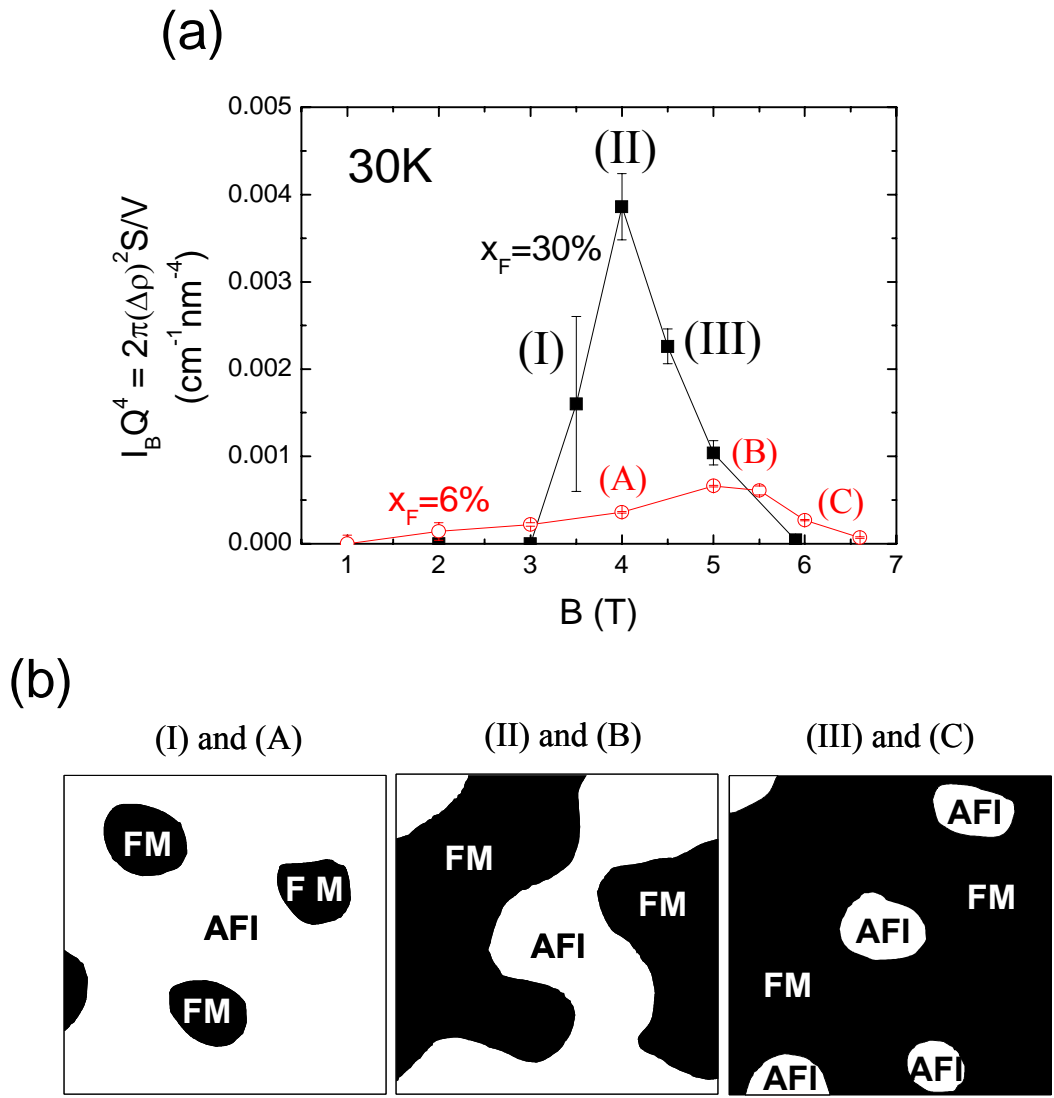

(a)

(b)

(I) and (A)    (II) and (B)    (III) and (C)

_Figure 5_

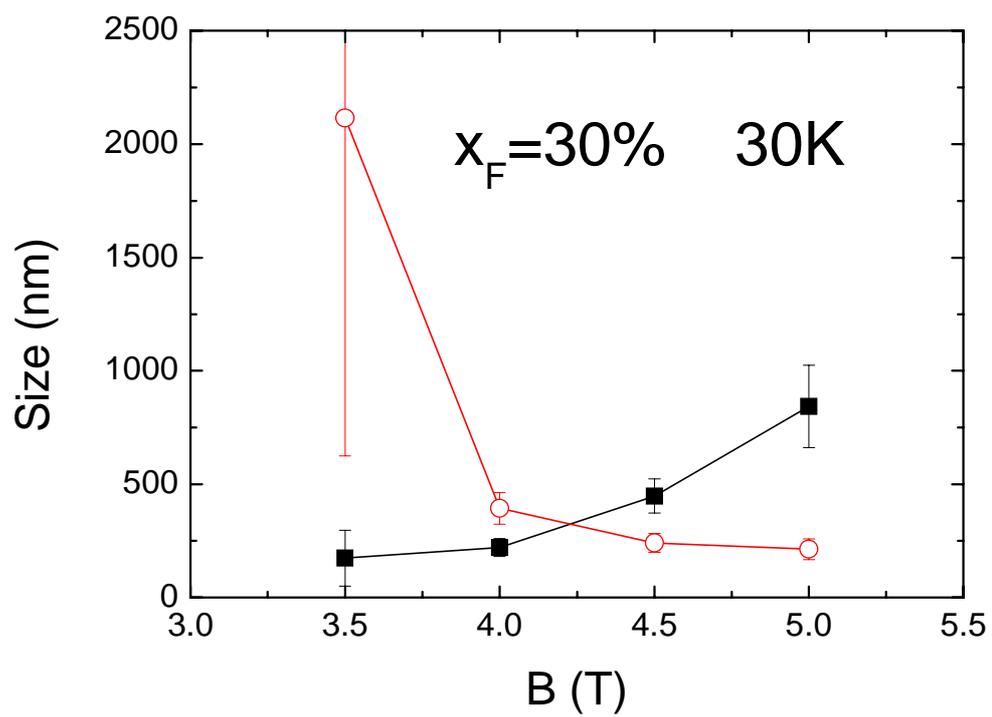

$x_F=30\%$    30K

*Figure 6*

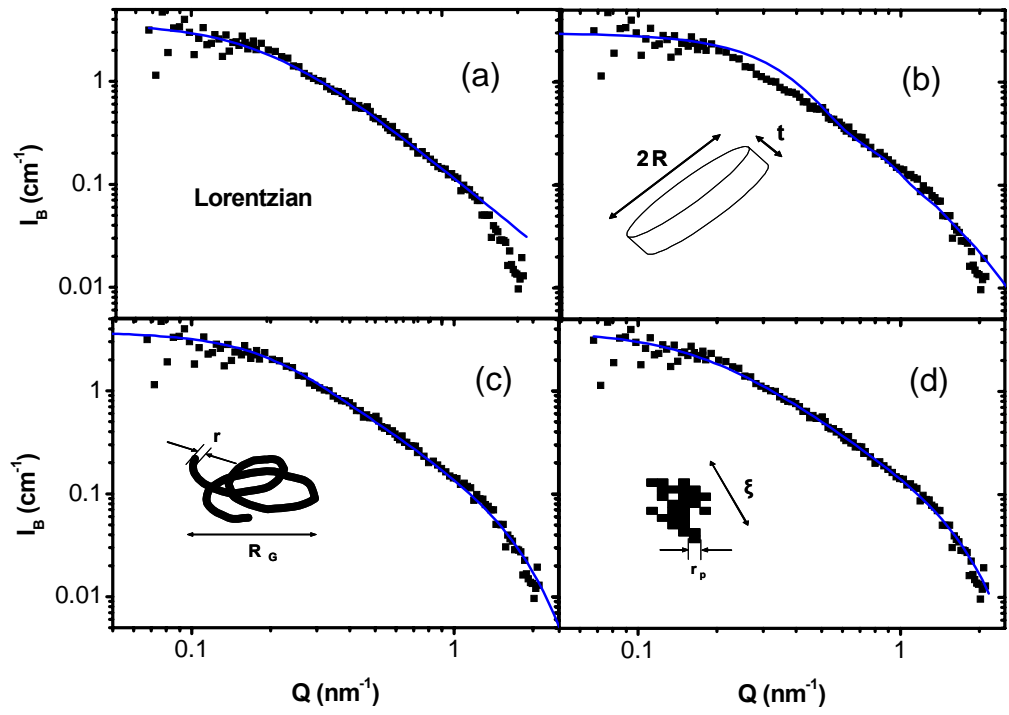

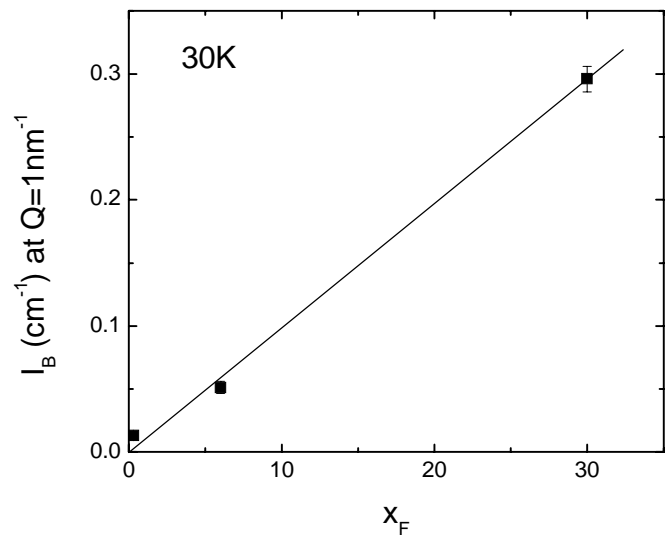

*Figure 8*

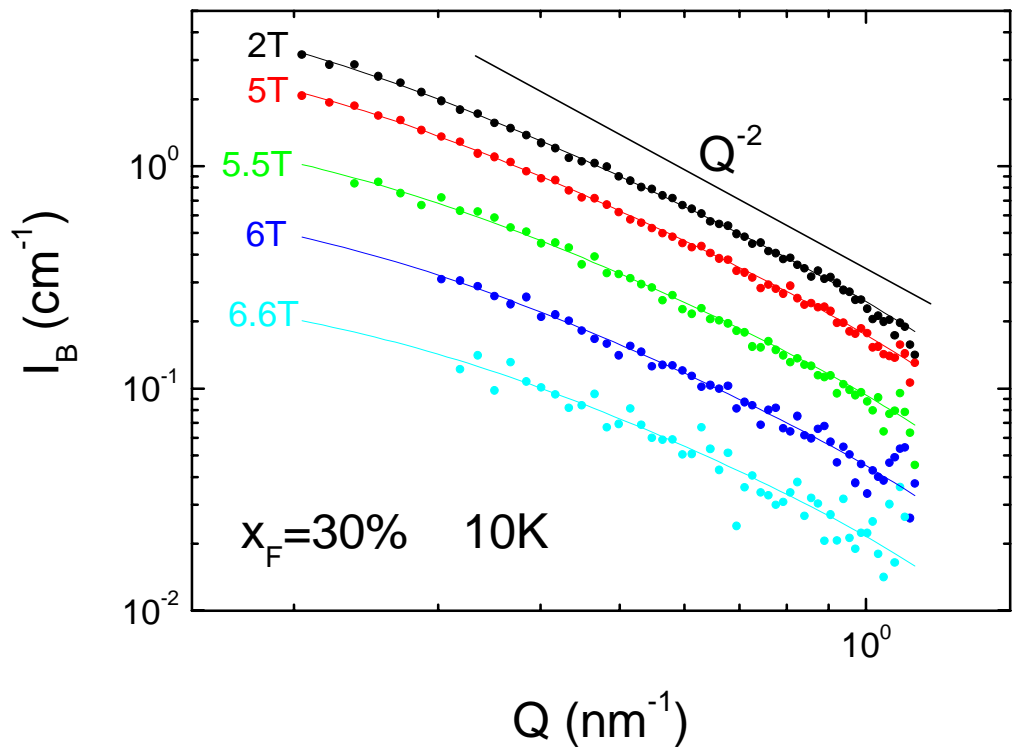

*Figure 9*

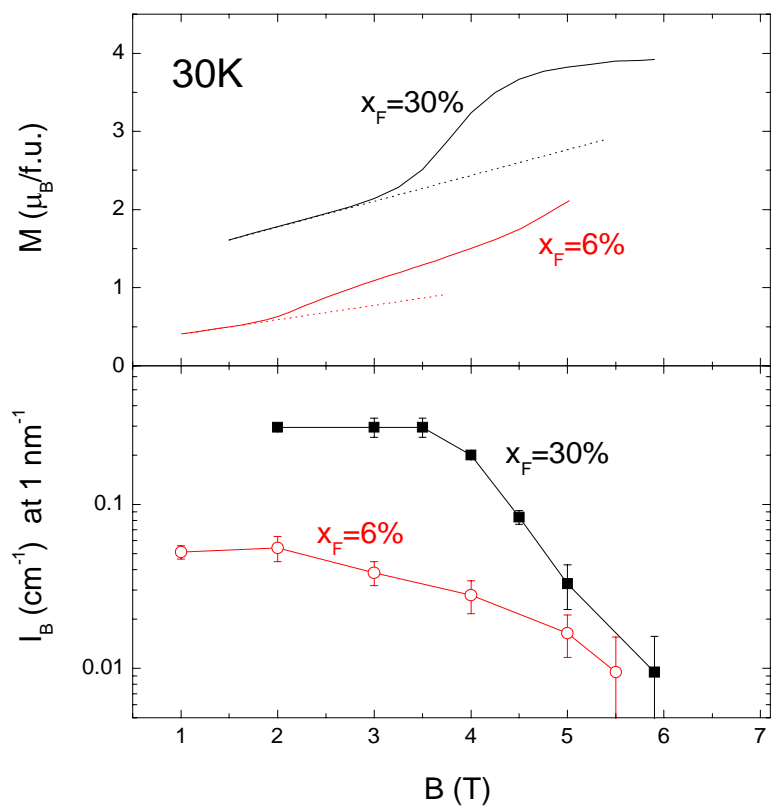

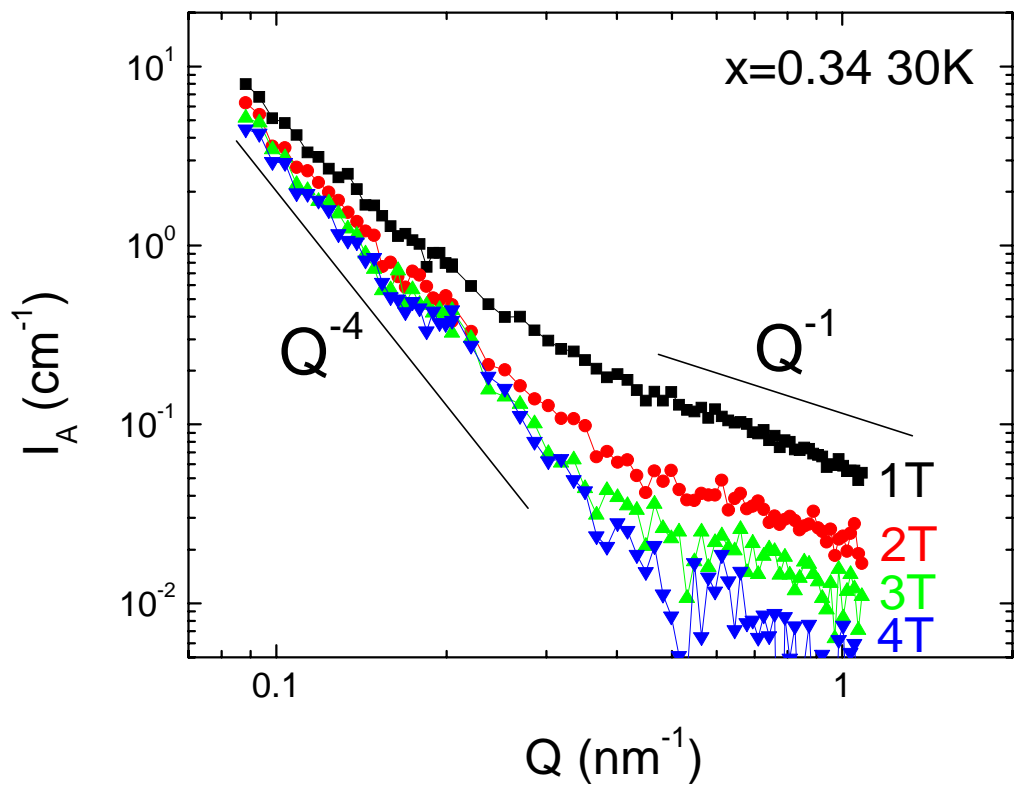

*Figure 11*

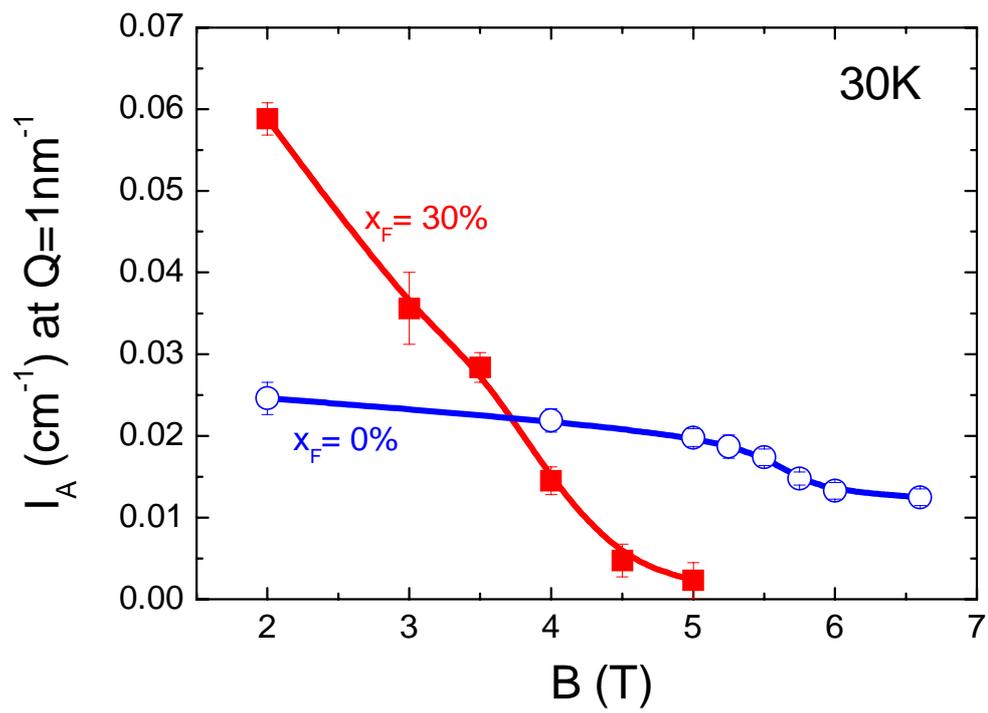